# The microstructural dependence of ionic transport in bi-continuous nanoporous metal

Congcheng Wang,[1,2,#] Anson Tsang,[1,3,#] Diwen Xiao,[1,2] Yuan Xu,[1,2] Shida Yang,[1,4] Ling-Zhi Liu,[5] Qiang Zheng,[6] Pan Liu,[6] Hai-Jun Jin,[5] Qing Chen[1,2,4,*]

[1]Department of Mechanical and Aerospace Engineering, the Hong Kong University of Science and Technology, Clear Water Bay, Kowloon, Hong Kong.
[2]The Energy Institute, the Hong Kong University of Science and Technology, Clear Water Bay, Kowloon, Hong Kong.
[3]George W. Woodruff School of Mechanical Engineering, Georgia Institute of Technology, Atlanta, GA, USA
[4]Department of Chemistry, the Hong Kong University of Science and Technology, Clear Water Bay, Kowloon, Hong Kong.
[5]Shenyang National Laboratory for Materials Science, Institute of Metal Research, Chinese Academy of Sciences, Shenyang, China.
[6]State Key Laboratory of Metal Matrix Composites, and Shanghai Key Laboratory of Advanced High-temperature Materials and Precision Forming, School of Materials Science and Engineering, Shanghai Jiao Tong University, Shanghai, China

[#]Equal contribution.
*Correspondence addressed to chenqing@ust.hk.

**Abstract**

Ionic transports in nanopores hold the key to unlocking the full potential of bi-continuous nanoporous (NP) metals as advanced electrodes in electrochemical devices. The precise control of the uniform NP metal structures also provides us a unique opportunity to understand how complex structures determine transports at nanoscales. For NP Au from the dealloying of a Ag-Au alloy, we can tune the pore size in the range of 13 nm to 2.4 µm and the porosity between 38% and 69% via isothermal coarsening. For NP Ag from the reduction-induced decomposition of AgCl, we can control additionally its structural hierarchy and pore orientation. We measure the effective ionic conductivities of 1 M $NaClO_4$ through these NP metals as membranes, which range from 7% to 44% of that of a free solution, corresponding to calculated pore tortuosity's between 2.7 and 1.3. The tortuosity of NP Au displays weak dependences on both the pore size and the porosity, consistent with the observed self-similarity in the coarsening, except for those of pores < 25 nm, which we consider deviating from the well-coarsened pore geometry. For NP Ag, the low tortuosity of the hierarchical structure can be explained with the Maxwell-Garnett equation and that of the oriented structure underlines the random orientation as the cause of slow transport in other NP metals. At last, we achieve high current densities of $CO_2$ reduction with these two low-tortuosity NP Ag's, demonstrating the significance of the structure-transport relationships for designing functional NP metals.

Pores at a nanoscale in solids give rise to ubiquitous transport phenomena whose importance transcends disciplines. The permeation of hydrocarbon through nanopores hampers the extraction of shale gas[1] but enables chemical separation with carbon, ceramic, and resin beds.[2] Ionic transports through charged narrow channels underpin the selectivity's of both biological[3] and synthetic membranes.[4] They are also crucial to electrochemical energy storage and conversion, where nanomaterials in the electrodes create pores at similar scales. Transports in these porous electrodes limit the performances of fuel cells,[5] batteries,[6] and supercapacitors.[7] Researchers have long strived to understand the microstructural dependence of transports so that they can design advanced electrodes to meet the ever-growing demand for high-power and energy-dense devices.

Among the emerging advanced electrodes are bi-continuous nanoporous (NP) metals. This class of materials, commonly fabricated via the selective dissolution of an alloy (*i.e.*, dealloying), comprise respectively continuous metal ligaments and pores both nanometer-wide.[8,9] The unique microstructures offers high electric conductivities, high specific areas full of exotic defects, and mechanical robustness, which underpin the successful applications as free-standing cathodes in fuel cells,[10] high-capacity anodes in metal batteries,[11,12] and high-power electron-collectors in pseudo-supercapacitors.[13]

Much work has suggested that the key to unlocking the potential of these materials lies with mass transports in the nanopores.[14–16] The most compelling evidence comes from the exploitation of structural hierarchy. The addition of another level of larger pores, intended as shortcuts for transports, improved the rate performance of NP Au as the cathode of a lithium-air battery,[14] the utilization of NP Ag as a Ag/AgO$_x$ cathode,[17] and the capacity of NP Ni in a pseudo-supercapacitor.[18] It would not have been the case if the transports in the nanopores were rapid enough. Other evidence includes a report of the negative effect of the small length scale of a NP Au skeleton on the accessible capacity of TiO$_2$ coating[19] and Lattice Boltzmann modelling by Falcucci *et al.* showing the shallow penetration of CO molecules into a NP Au catalyst,[16] later supported by experiments.[20]

Yet, it remains elusive how sluggish the transports are and how they depend on the structures. Among few attempts to quantify the transports, Seker *et al.* measured the diffusivity of a drug-surrogate through NP Au film to be ~50% lower than that in a free solution.[21] Both Ding *et al.*[10] and Biener *et al.*[20] measured gas permeability through NP Au and showed that the structure significantly impeded the flow. Bailey *et al.* demonstrated the influence of NP

Au surface on the diffusivity of molecules via an applied voltage.[22,23] Recently, Haensch *et al.*[24] used scanning electrochemical microscopy to quantify the diffusivity of ascorbic acid in NP Au to be 50 – 67% of that in a free solution, but the results were derived through numerical simulation with several fitting parameters. Lu *et al.*[25] calculated tortuosity factors as large as 20 for the Knudsen diffusion of metal atoms leaving NP metal via modelling the kinetics of vapor phase dealloying, which is nonetheless different from electrochemical dealloying in terms of the resulted structures. No insight into the structural dependence of the transports has emerged.

Here we give the first account for the relationship between the rates of ionic transport and the structural characteristics of NP metals. By measuring the impedance through NP metal as a membrane, we determine the effective ionic conductivities in the pores of NP metals to range from 7% to 44% of those of free electrolytes. We analyze the results in terms of the pore tortuosity and its dependence on the pore size and the porosity. A strong pore-size dependence emerges below a size threshold ~25 nm, and a strong porosity-dependence shows up for NP Ag fabricated via reduction-induced decomposition but not for NP Au fabricated via dealloying. Underlying the structural dependence is the variation in the pore geometry. At last, we demonstrate the practical relevance of the work through the application of NP Ag in $CO_2$ reduction.

**Fabricating NP metals**

We focus on two types of NP metals and two of their structural characteristics. The first type is NP Au from the dealloying of a Ag-Au alloy, the model structure whose morphology has been extensively studied.[26] The other is NP Ag fabricated by reduction-induced decomposition (RID).[17] This dealloying-analogy, which selectively dissolves anions from an otherwise insoluble compound precursor, was developed to broaden the spectrum of achievable NP structures. We have demonstrated the control over the porosity,[17,27] the structural hierarchy,[17] and the structural orientation[28] through RID. Unlike NP Au from dealloying, NP Ag's from RID represent those tunable for functional needs.

The two structural characteristics are the average pore size (*d*) and the porosity (ε). *d* corresponds to the average width of pore channels, estimated with the open-source code AQUAMI.[29] ε corresponds to the volume fraction of pores accessible by a fluid, or commonly referred to as open porosity. Though rarely discussed, closed pores do exist inside the ligaments due to the Rayleigh-Plateau instability which pinches off the nanopores.[30,31] We thus first measure the dimensions of a sample, derive the density of the metal phase based on the composition if it is an alloy, and thereby calculate the sum of open and closed porosities. We then use a pycnometer (Fig. S1) to estimate and subtract out the volume of closed pores.

Figures 1a and b show the typical cross-sectional morphology of NP Au under a scanning electron microscope (SEM). It was fabricated by dealloying a 250 µm thick, 0.5 cm$^2$ large sheet of $Ag_{0.8}Au_{0.2}$ in 1 M perchloric acid at relatively low potential to retain a fraction of Ag for ε close to that of NP Ag. $d$ is 13 nm with a relatively narrow distribution (the inset of Fig. 1a). ε is 68.9%, with <0.5% of closed pores as in all other NP Au samples, consistent with previous work.[31] This structure will be referred to as the base case of NP Au. $d$ increases slightly to 15, 18, and 21 nm (Fig. S2 a to c) with <1% reduction in ε when coarsened at 50, 100, and 150 ºC for 2 hours, respectively. Coarsening above 200 ºC leads to densification. At 600 ºC, ε was reduced to 47% (Fig. 1c) and at 800 ºC to 38% (Fig. S2 d to f).

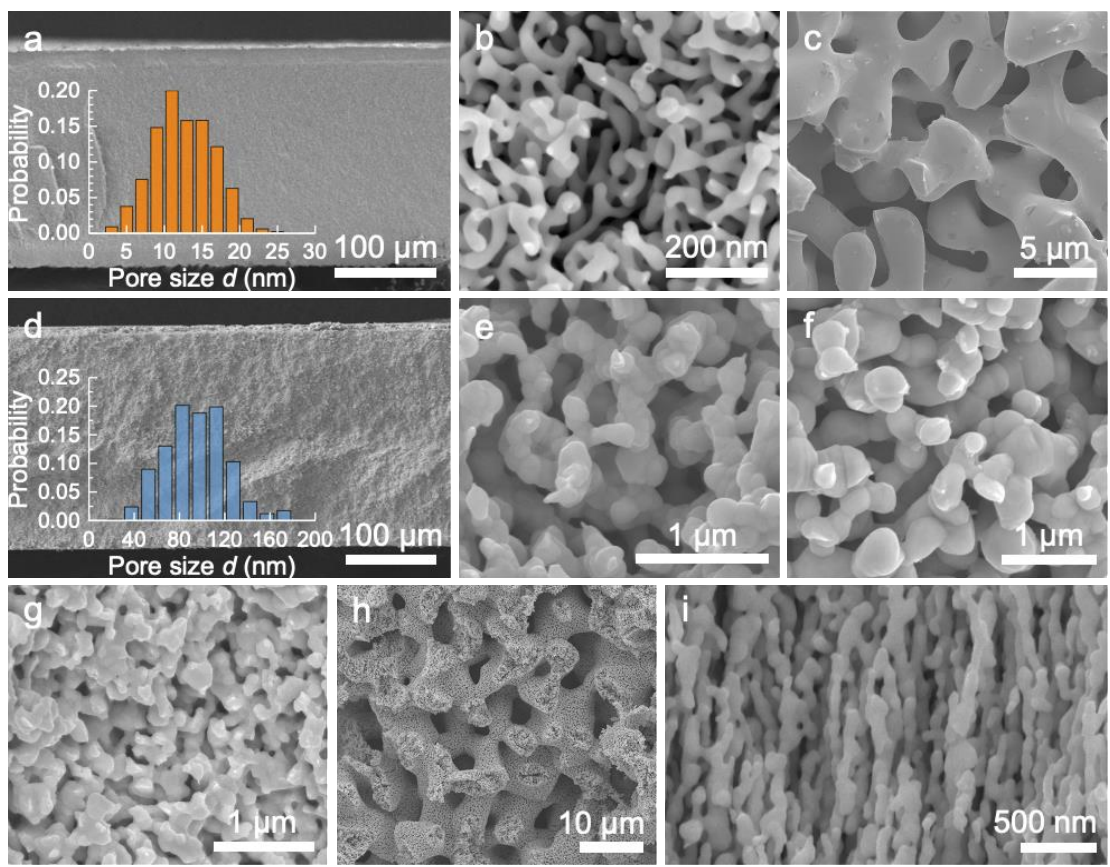

**Figure 1.** Morphologies of NP Au and Ag captured under SEM. (a) and (b) the base-case NP Au at two different magnifications. The inset of (a) shows the pore size distribution. (c) NP Au annealed at 600 ºC for 2 hours. (d) and (e) the base-case NP Ag at two different

magnifications. The inset of (d) shows the pore size distribution. (f) NP Ag annealed at 200 ºC for 1 hour. (g) Low-ε NP Ag fabricated from $Ag_2O$. (h) Hierarchical NP Ag. (i) Oriented NP Ag. All images were taken at the cross-sections.

Figures 1d and e show the typical morphology of NP Ag fabricated through RID of AgCl. The reduction was carried out with a solution of $VSO_4$ and $H_2SO_4$ instead of $NaBH_4$ to avoid hydrogen bubbles. With $d$ = 95 nm and ε = 61%, the structure, referred to as the base case of NP Ag, resembles NP Au but with important differences. It has a higher fraction of closed pores (1.6%) and an average grain size close to the width of the ligaments, whereas NP Au inherits the hundred micron-sized grains from the alloy precursor. The small grains are noticeable even under SEM for their grooves on the surface. At 200 ºC for 1 hour, $d$ increases to 499 nm, ε stays at 58%, and the grooves deepen (Fig. 1f). Further raising the temperature leads to a loss of the structural integrity likely due to the more rapid pinch-off at the grain boundaries, which may have also led to the broadening of pore size distribution as it coarsens (the standard deviation can be found in Table S1), despite the initial distribution (the inset of Fig. 1d) similar with that of NP Au.

With RID, we can fabricate a variety of other NP Ag structures with different precursors.[27,28] NP Ag with low ε (37%, Fig. 1g) can be achieved by using $Ag_2O$ instead of AgCl.[27] Hierarchical structures (Fig. 1h and S2h) can be formed with porous precursors made of either AgCl and $Ag_2O$ respectively via powder sintering. RID of the porous AgCl precursor leads to an upper-level pore size ~5 µm and ε of 78% (Fig. 1h). As $Ag_2O$ is unstable in an acid, all the above three samples were reduced with $NaBH_4$, which leads to smaller $d$ (65 – 68 nm). At last, we can align the pores along the direction of the RID reaction. When we flow the solution of $VSO_4$ in a flow cell, the $VSO_4$ concentration at the reaction front can be maintained at a high value to render rapid RID. Virtually all pores are oriented through the plane of the NP Ag sheet (Fig. 1i). ε is 54% in this case.

**Measuring effective conductivity**

The measurement of ionic transport, illustrated in Fig. 2a, is designed with the following considerations. We use NP metal as a membrane instead of an electrode to remove the complication from electron transfer and electron conduction. A typical measurement of ionic conductivity through a membrane would involve four electrode probes, two far apart undergoing a diffusion-controlled electrochemical reaction and two touching the two sides of the membrane to sense the liquid potential.[32] However, the electric conductivity of NP metal forbids its contact with probes, so we design a jig (Fig. 2a and S3) to place two Pt probes very close (920 µm) to NP metal to collect electrochemical impedance spectroscopy (EIS). EIS, instead of direct current, is used so that we can investigate a concentrated electrolyte of a pair of weakly adsorbing small ions, $Na^+$ and $ClO_4^-$ to facilitate a comparison between Au and Ag and generalize conclusions to other metals. Direct current would have required two Ag/AgCl probes and a chloride-electrolyte,[22] which can coarsen NP Au and corrode NP Ag.

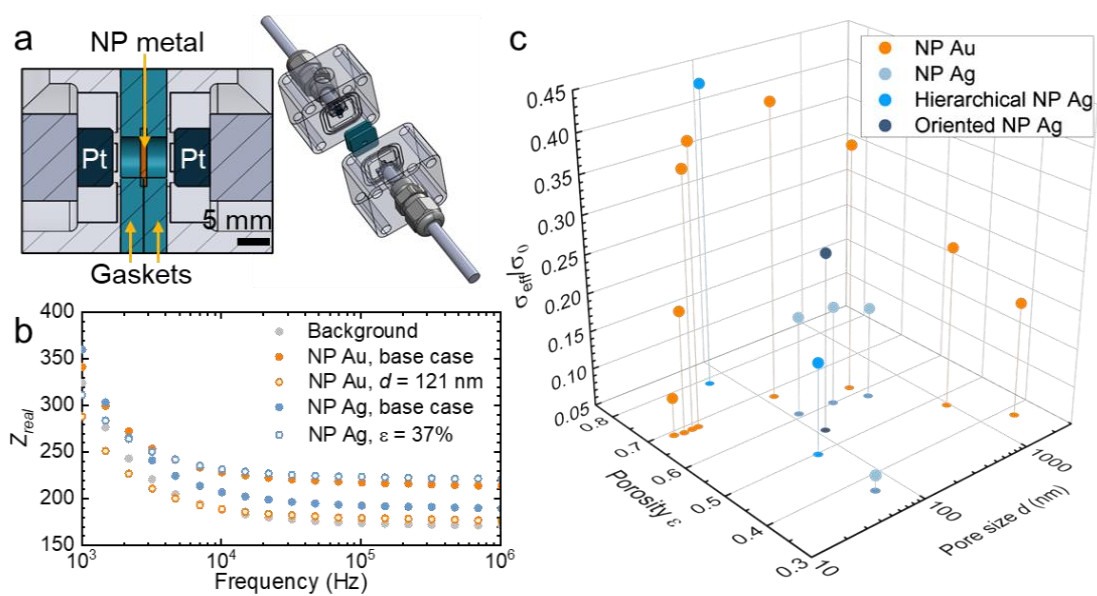

**Figure 2.** The measurement of effective ionic conductivity through NP metal. (a) The design of the measurement jig with key components labelled. On the left, we highlight the electrolyte chamber. (b) Representative EIS results in a plot of $Z_{real}$ vs. frequency. (c) All the measured values of $\sigma_{eff}/\sigma_0$. We project all the datapoints to the bottom plane to facilitate the reading of $\varepsilon$, $d$, and $\sigma_{eff}/\sigma_0$, the last of which corresponds to the length of the line between the point and its projection. For the hierarchical structures, $d$ is of the lower-level smaller pores. The color code, orange for NP Au and different shades of blue for different types of NP Ag, remains the same throughout the work.

A representative set of EIS curves for calculating the effective conductivity are shown in Fig. 2b. The grey curve corresponds to a background measurement, performed with the assembled jig with 1 M $NaClO_4$ but without NP metal. The real component of the impedance ($Z_{real}$), dictated by the ionic resistance of the electrolyte between the two probes, is nearly constant at high frequencies. The placement of NP metal in the jig shifts the curve upwards and increases the resistance. As the overall distance between the two probes is kept the same with two highly compressible thick gaskets, we can use the difference between the resistances with and without NP metal (ΔR) in the frequency range of $10^5 - 10^6$ Hz to calculate a ratio between the effective conductivity through the pores ($\sigma_{eff}$) and the conductivity of a free solution ($\sigma_0$) with

$$\frac{\sigma_{eff}}{\sigma_0} = 1/(1 + \frac{\sigma_0 A \Delta R}{l}), \qquad (1)$$

where $A$ is the area of the NP metal not covered by the gaskets (0.07 cm$^2$) and $l$ the thickness of the NP metal determined with SEM. The measurement results are summarized in Table S1.

In Fig. 2c, we plot the values of $\sigma_{eff}/\sigma_0$ against $d$ and $\varepsilon$. They fall in a wide range of 0.07 – 0.44. The conductivity increases generally with increasing $d$ and $\varepsilon$. NP Au with small $d$ and NP Ag with low $\varepsilon$ both slow down the rate of ionic transport by more than ten times, whereas NP Au with medium $d$ and hierarchical NP Ag retain > 40% of the conductivity. With similar $d$ and $\varepsilon$, NP Au offers higher $\sigma_{eff}$. The diverse NP Ag structures lead to a large variety in $\sigma_{eff}$, though roughly following a positive correlation with $\varepsilon$. The results promise the resolution of quantitative structure-property relationships upon further analyses.

**The pore-size effect**

Our analysis is centered around pore tortuosity ($\tau$), defined with the following equation,

$$\frac{\sigma_{eff}}{\sigma_0} = \frac{\varepsilon}{\tau^2}. \qquad (2)$$

$\varepsilon$ in the numerator, taken as the areal fraction of the pores, corrects the transport property to be specific to the pores. $\tau$ in the denominator, as the name infers, pertains to the tortuous pathlength of the ions through the NP metal. Though derived for capillary pores, the proportionality of $\sigma_{eff}/\sigma_0$ to $\tau^2$ is widely accepted for the consistent dimensions.[6,33,34] A porous sheet made by punching straight, large and separated cylindrical holes would have $\tau = 1$ and $\sigma_{eff}/\sigma_0$ the same as $\varepsilon$. A change from the cylindrical pores to a more complex geometry increases $\tau$ above 1. So does a sufficiently small pore size. Although the use of $\tau$ has long been criticized as being empirical,[34,35] we choose it for two reasons. First, its interpretation can be rigorous for a well-defined pore geometry.[33,36] Second, there are many $\tau$ values to compare our measurements with. They include experiments on a variety of porous media[6,37,38] and simulations based on tomographic images[39] that have achieved a fair agreement with experiments.

In Fig. 3a, we gauge the effect of pore size by including $\tau$ values calculated for those of different $d$'s but similar $\varepsilon$'s (68 – 69% for NP Au and 58 – 61% for NP Ag). At $d > 25$ nm, $\tau$ is stable against more than a decade of length scale change for both NP Au and Ag. The lowest values are 1.30 for NP Au and 1.85 for Ag, respectively (highlighted with dashed lines). At $d < 25$ nm, $\tau$ of NP Au shows a strong dependence on $d$. A size effect at this scale is remarkable. Studies on nanochannels in silica,[40] multiwall carbon nanotubes,[41] and cylindrical pores drilled in mica[32] have all suggested no difference between $\sigma_{eff}$ and $\sigma_0$ for concentrated electrolytes in pores as small as 3.4 nm. Models for more complex porous media,[42] though applied primarily to molecular diffusion in polymer membranes and silica particles, would predict similarly no size effect as the hydration radii of $Na^+$ and $ClO_4^-$ (~0.35 nm[43]) are much smaller than even the smallest $d$ examined here.

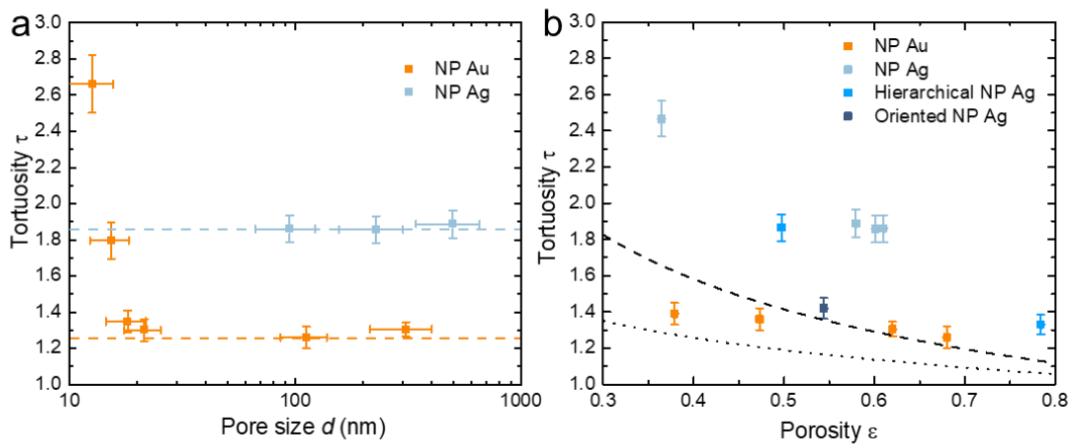

**Figure 3.** The dependence of $\tau$ on (a) $d$ and (b) $\varepsilon$. Only samples of $\varepsilon$ in the range of 68 – 69% and 58 – 61% for NP Au and Ag respectively are included in (a), and those of $d > 25$ nm in (b). The lowest values of $\tau$ are highlighted with dashed lines in (a), and the predictions by the two Bruggeman equations are plotted in dashed and dotted lines in (b).

We propose an explanation based on the pore geometry. It is inspired by three observations. First, NP Au seems to coarsen slowly below 150 ºC where $d$ stays below 25 nm. The size increases drastically to 112 nm at 200 ºC, an abrupt change in the kinetics more evident in a plot of $ln(d)$ vs. the reciprocal of temperature (Fig. S4), the different slopes of which below and above 200 °C suggest different power laws and/or activation energies for coarsening.[17,44] Second, in the base-case NP Au, we find an appreciable number of extremely slim ligaments with a very high aspect ratio (Fig. 4a) not easily spotted in NP Au with $d > 18$ nm. Third, the threshold ~25 nm coincides with the steady-state length scale expected for NP Au for a long period of coarsening in a weakly-adsorbing electrolyte at room temperature.[45] We speculate that a mechanism, yet to be identified but crucial to the initial slow coarsening, prevents slim ligaments from being pinched off, sustaining high connectivity among the ligaments and thus low connectivity in the complementary pores. Further coarsening brings the ligament connectivity down to a stable value, and τ thereby plateaus. Such a connectivity change, usually characterized by the scaled genus, has been observed or implied in the literature, though not yet in a quantitative agreement with our results.[31,46] The size effect is thus not caused directly by a change in the pore size but rather by that in the pore geometry as the structure coarsens.

We corroborate the explanation with a measurement in an aqueous solution of 0.5 M zinc acetate. This pair of ions diffuse at much slower rates than $Na^+$ and $ClO_4^-$ due to larger hydration radii.[47] We expect a different interaction with the Au surface, too. Hence, if the size effect was from either the respective size of the pores and the ions or the specific interactions between the ions and Au, we would measure a lower value of τ with zinc acetate. Instead, for NP Au with $d = 18$ nm (chosen for its higher stability in different electrolytes), we get τ = 1.79 (Fig. S5), whose difference from that with $NaClO_4$ is within the error. It supports our explanation based on the geometry, which should be independent of the electrolyte.

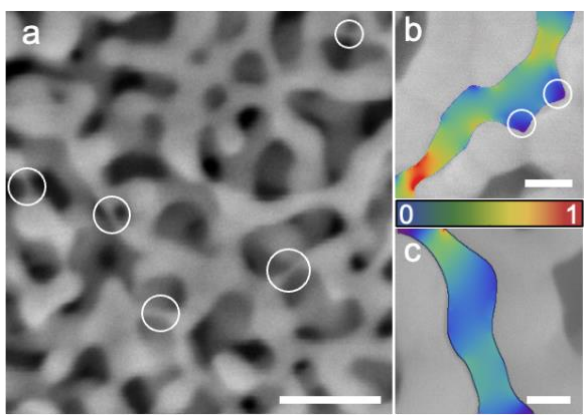

**Figure 4.** Geometric interpretations of the measured values of τ. (a) SEM of the surface of the base-case NP Au, where we highlight the slim ligaments with white circles. (b) and (c) The simulated distributions of normalized flux in the pores of the base-case NP Ag and NP

Au with $d$ = 112 nm, respectively, overlaid on the original SEM images used for the simulation. The flux is colored according to the scheme in (b). All scale bars are 50 nm.

**The porosity effect**

We place $\tau$ values for all the NP metals with $d > 25$ nm on the same plot to gauge the impact of $\varepsilon$ (Fig. 3b), where we find major differences between NP Au and Ag. The decrease in $\varepsilon$ via coarsening leads to a very limited increase in $\tau$ for NP Au. NP Ag instead displays a monatomic increase in $\tau$ as $\varepsilon$ decreases, except for that of oriented pores. With similar $\varepsilon$, NP Ag has higher $\tau$ than NP Au, which explains the difference in Fig. 3a. Neither NP Au nor Ag follows the popular Bruggeman equation that assumes either randomly packed spheres ($\tau^2 = \varepsilon^{0.5}$, the dotted line in Fig. 3b) or cylinders ($\tau^2 = \varepsilon$, the dashed line),[48] hardly a surprise as neither geometry resembles NP metal.

We can identify two features in NP Ag responsible for the large $\tau$. The first is the less uniform pore size, evident in the distribution of pore size (Table S1). The second is the large number of grain boundary grooves. We can visualize their impacts by simulating random walk through a nanopore extracted from an SEM image. The simulation uses an open-source code, TauFactor,[39] and imposes a concentration different between the top and bottom boundaries of the image. In Fig. 4b, we overlay the simulated, normalized flux with the original SEM image. The narrower end of the pore in NP Ag limits the outgoing flux, an effect described in literature usually as constrictivity due to the variation in the pore width.[34,38] Regions near the two deep grooves (highlighted by circles), dead ends in two dimensions and detours in three, see little flux. In contrast, a typical pore in NP Au of a similar pore size displays a more uniform distribution (Fig. 4c).

The stable $\tau$ of NP Au against the drastic changes in $\varepsilon$ (69% to 38%) and $d$ (21 to 2447 nm) adds new evidence to the self-similarity of the isothermal coarsening.[49,50] Unlike previous studies focused on topology and solid mechanics, our data speaks from the pore geometry seen by ions. The most remarkable observation is the stable pore geometry against the substantial densification. It agrees with the conclusion by a previous study that the densification retains the network topology.[51] As both the pores and the ligaments grow in size, the topology stays by reducing the number of pore channels per volume, whose impact on the transport has been accounted by normalizing $\sigma_{eff}/\sigma_0$ with $\varepsilon$ (as in Eq. 2), so $\tau$ stays constant. For NP Ag, although the coarsening does not change $\tau$ either, we can only claim the self-similarity concerning changes in the pore size but not changes in the porosity, given the limitation in tuning porosity via the coarsening.

Underlying the strong ε-dependence for NP Ag is the variation among structures fabricated with the different precursors. Without a more quantitative measure of the structural difference, we can only attribute the high τ of the low-ε NP Ag to the higher connectivity among the stouter ligaments formed by RID of $Ag_2O$, apparent when comparing Fig. 1e and g. The impact of the structural hierarchy can be better explained. Were the upper-level porosity of the same geometry as the lower level, we would expect the same τ for those with and without the hierarchy, given the independence on the pore size. Yet, the two levels are deemed different by the fabrication method. The upper-level porosity is formed by sintering particles of a relatively uniform size, closer to an agglomerate whose effective property can be calculated with the Maxwell-Garnett formula.[52] To apply this formula, we consider the hierarchical structure as the random inclusion of NP Ag particles into the bounding volume of the whole structure. Assume the particles of the same porosity and effective conductivity as that formed by RID of the dense, non-porous precursor (termed as $\varepsilon_0$ and $\sigma_{eff,0}$). The volume occupied by the NP Ag particles is $(1 - \varepsilon)/(1 - \varepsilon_0)$. $\sigma_{eff}$ of the hierarchical structure can then be calculated via

$$\sigma_{eff} = \sigma_0 \left\{ 1 + \left[ 3\left(\frac{\sigma_{eff,0}}{\sigma_0} - 1\right)\frac{1-\varepsilon}{1-\varepsilon_0} \right] / \left[ \frac{\sigma_{eff,0}}{\sigma_0} + 2 - \left(\frac{\sigma_{eff,0}}{\sigma_0} - 1\right)\frac{1-\varepsilon}{1-\varepsilon_0} \right] \right\}. \tag{3}$$

The as-calculated values of the hierarchical structures from the porous AgCl and $Ag_2O$ precursors are 0.47 and 0.21, respectively, very close to the measured value of 0.44 and 0.17. Behind the good agreement is the root benefit of the structural hierarchy, the less tortuous upper-level pore geometry built via powder sintering.

We can further compare our measurements to values of porous electrodes from the literature. Most of them have ε < 50% given the common need for a high volumetric capacity or area. For those of ε ~50%, τ values from electrochemical measurements and flux-simulation of tomographic images fall in a broad range of 1.2 – 2.7.[6,33,37] At the lower ends, we have electrodes made from binding $LiFePO_4$ particles or sintering Ni–yttria stabilized zirconia.[37,53] Given the higher structural uniformity, we would expect even lower τ from NP Au. Yet, the high connectivity among the ligaments, difficult to achieve particularly with a particle assembly, may have rendered τ of NP Au in the same range. A similar argument can be made for the relatively high τ of NP Ag.

At last, the low τ of the oriented NP Ag (dark blue in Fig. 3b) underlines the unsurprising source of tortuosity in other NP metals, the random pore orientation. Pores aligned away from the direction of transport account for a large fraction of the porosity but a limited contribution to the transport. τ of the oriented NP Ag is low but not yet close to unity, as the pores do not go all the way through the sheet due to the formation mechanism[28] and pinch offs; otherwise, the sample would have lost its integrity. Nonetheless, the control over the orientation in RID adds a new dimension to the fabrication of NP metal for accelerating transport.

**Catalyzing $CO_2$ reduction with improved transports**

We test three samples of NP Ag of distinct transport properties as free-standing electrodes for the reduction of $CO_2$ to CO. We choose this reaction not only for its practical significance but also for its strong dependence on the thickness of a three-dimensional porous electrode, whose area can only be effectively utilized at high rates of mass transport through the pores. We fabricated the electrodes all ~100 µm thick, beyond the typical depth reachable by the reactants to showcase the impact of transport.[54] The reaction was performed in a three-electrode cell of a $CO_2$-saturated $KHCO_3$ solution, without the direct involvement of $CO_2$ gas to avoid triple-phase boundaries beyond the implication of the investigated transport properties. At each voltage, the rate of CO generation was recorded for one hour with online gas chromatography and converted via Faraday's law to a current density ($i_{CO}$) normalized to the projected area. $i_{CO}$ is divided by the overall current density to afford a Faradaic efficiency.

The oriented and the hierarchical NP Ag electrodes achieve substantially higher values of $i_{CO}$, as shown in Fig. 5. All three deliver $i_{CO}$'s >10 mA/cm$^2$ and Faradaic efficiencies over 98% at -0.7 V vs. the standard hydrogen electrode (SHE), among the highest reported for monolithic Ag electrodes and consistent with the high catalytic activity achieved by Ag nanostructures fabricated via methods similar to RID[55–57]. At this high over-voltage where the limit by mass transport should be significant, $i_{CO}$'s on the oriented and the hierarchical electrodes are 45% and 28% higher than that on the base case, respectively, despite the similar electrochemical surface areas determined by the underpotential deposition of Pb (Fig. S6). The differences arise from the amount of accessed surface area deep in the electrodes, which can be approximated with the diffusional length of the rate-limiting species. Assume its effective diffusivity $D_{eff}$ normalized by the free-solution diffusivity to be the same as $\sigma_{eff}/\sigma_0$. As the diffusional length is proportional to the square root of $D_{eff}$, based on the values of $\sigma_{eff}/\sigma_0$, we would predict 56% and 25% higher $i_{CO}$'s with the hierarchical and oriented structures, respectively, than the base case, in a good agreement with our measurements. The hierarchical NP Ag seems to underperform. A possible explanation is the unfavorable dynamics of CO on its surface, creating visible bubbles and likely responsible for the large errors in its efficiency, too.

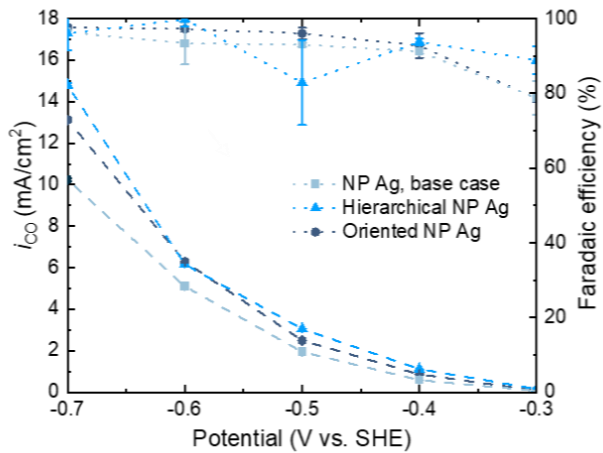

**Figure 5**. A comparison of three NP Ag's for the electrocatalysis of $CO_2$ reduction. Values of $i_{CO}$ (the left axis) are connected by dashed lines and Faradaic efficiencies (the right axis) by dotted lines.

**Discussion**

The observed structure-transport relationships should be general for diffusion or migration of small ions and molecules in a liquid solution through NP metal. They are all attributed to differences in pore geometry, which are independent of the size or composition of either the ion/molecule or the metal. Exceptions will occur when surface transports become significant in extremely small pores and dilute electrolytes. We recommend the $\tau$ values of NP Au and Ag as first estimates of lower and upper bounds respectively for any NP metal of a similar pore size and porosity, as we consider most NP metals in the literature between them in terms of the structural uniformity, surface roughness, and pore geometry, among other influential characteristics we have not quantified. These bounds, nonetheless, enclose a rather wide range of transport rates, which underscore the need for the structure-transport relationships in the design of functional NP metal.

Two of the key relationships we have revealed pertain to two common design strategies. The first relates to the pore size. While much work has implied slow transport in NP metal because of the small pore size, our results speak against it for pores roughly above 30 nm and suggest its unlikelihood even down to ~15 nm, consistent with the literature on nanofluidics[40,41] and other porous media.[42] There is thus no apparent dilemma between a high surface area and rapid transport. Yet, pore size can impact the transport through the pore geometry, whose variation among NP metals is often noticeable in the morphologies. The second popular strategy is structural hierarchy. Instead of the large pores, its impact on transports comes from two factors. The first is high $\varepsilon$, whose importance has not been stressed enough for NP metal. Even without hierarchy, the effective rate of transport can increase linearly with $\varepsilon$ (as in Eq. 2). Yet without it, NP metal would easily lose its integrity due to the Rayleigh-Plateau stability.[58] Structural hierarchy is thus a means towards high-$\varepsilon$ NP metal. The second factor is the pore geometry of the upper-level porosity, which can facilitate transport while retaining the morphology of NP metal desired for the functionality.

At last, there are immense opportunities of developing fundamental insights into both NP metal and transport phenomena at nanoscales by marrying the two. For NP metal, not only does our work help improves the functionality, it also reveals a different facet of the pore geometry. The geometry, particularly the topology of NP metal, has been at the center of recent discussions of the ubiquitous mechanical properties.[51,58,59] For studying transports at nanoscales, NP metal is scalable in all dimensions, uniform and tunable in its structural characteristics, and stable in different fluid environments. The metallic nature has allowed in-depth characterizations of its structure down to atomic scales,[26,60] and the deep understanding of its formation mechanism has led to faithful structural recreations via mathematical methods.[50] NP metal holds a promise of a model structure to bridge the gap between theories developed for single or few nanopores and the empirical models for complex porous materials.

**Methods**

*Fabrication:* NP Au samples were prepared by electrochemical dealloying of a $Ag_{0.8}Au_{0.2}$ alloy. The dealloying was carried out in a 1 M $HClO_4$ (VWR) solution by applying a fixed voltage at 0.75 V vs. a Ag/AgCl reference electrode for 24 h. NP Ag samples were fabricated by RID of AgCl or $Ag_2O$ precursors. AgCl precursors were prepared by melting or pressing

AgCl powders synthesized by mixing solutions of 0.5 M AgNO$_3$ (Sigma Aldrich) and 1 M NaCl (VWR). Ag$_2$O precursors were prepared by pressing Ag$_2$O powders synthesized by mixing solutions of 0.5 M AgNO$_3$ and 1 M NaOH (Sigma Aldrich). The melting process of AgCl was carried out in a quartz crucible at 500 ℃ for 15 min in air. AgCl or Ag$_2$O powders were pressed into pellets with a hydraulic press machine. RID experiments were performed by immersing the compound precursors into 1 M VSO$_4$ (Sigma Aldrich) in 1 M H$_2$SO$_4$ (VWR) solutions or 0.1 M NaBH$_4$ (Sigma Aldrich) solutions. Oriented NP Ag was fabricated via RID in a flow cell with a 1 M VSO$_4$ in 1 M H$_2$SO$_4$ solution as the flowing electrolyte. Annealing experiments were carried out in Ar in a tube-furnace. All dealloyed and decomposed samples were thoroughly washed with deionized water and dried in vacuum before measurements and characterization.

*Structural characterization:* Scanning Electron Microscopy (SEM) was conducted with either a JEOL-7100F scanning electron microscope or an FEI Helios G4 UX FIB/FESEM system, which contains both a focused ion beam and an ultra-high resolution field emission scanning electron microscope. Energy dispersive spectroscopy was performed in the same microscope to determine the composition of NP metal.

*Conductivity measurement:* Conductivity measurement was carried out in an acrylic jig shown in Fig. 2a and Fig. S4. Before the measurement, a sample was immersed in 1 M sodium perchlorate (Sigma Aldrich) or 0.5 M zinc acetate (Sigma Aldrich) under a low vacuum of 0.1 Bar for 3 min to remove the air inside and make sure that the pores were filled with the electrolyte. The measurement cell was then assembled and put under the same vacuum condition for another 3 min before measurement. EIS was carried out with a Bio-logic VMP3 potentio-stat, with a frequency range from 100 Hz to 1 MHz and an AC amplitude of 10 mV.

*Image analysis and simulation:* An image was first binarized under a threshold that renders the fraction of black pixels (the pores) the same as the porosity of the corresponding NP metal. The binary image was then put into AQUAMI (https://pypi.org/project/aquami/) to estimate the pore size and TauFactor (https://sourceforge.net/projects/taufactor/) to simulate random walk, respectively.

*CO$_2$ reduction:* A gastight H-type electrochemical cell separated by a Nafion-117 proton exchange membrane was used to evaluate the catalytic performance of nanoporous Ag for CO$_2$ reduction. Each chamber was filled with 40 ml of 0.5 M KHCO$_3$ (99.7%, Sigma

Aldrich) solution (pH=7.3) with ~10 ml headspace. The KHCO$_3$ was used without purification. The working, counter, and reference electrodes were nanoporous Ag, Pt mesh, and Ag/AgCl electrode, respectively. The potential was automatically compensated with an 80% level of IR compensation. Before each constant potential electrocatalysis, the cathodic chamber was purged with CO$_2$ (99.999%, Scientific Gas Engineering Co., LTD) for at least 1 hour to remove the residual air. Continuous CO$_2$ was purged into the cell at a rate of 30 sccm by a mass-flow controller (Sevenstar, Beijing). The electrolyte was stirred at 700 rpm by a magnetic stirrer. Effluent gas from the cathodic chamber was fed to an online gas chromatography (GC2060, Ramiin, Shanghai). A flame ionization detector and a thermal conductivity detector (TCD) were used for H$_2$ and CO detections, respectively. Gas chromatography automatically records at the 15th, 28th, 41st, and 54th min of the one-hour electrolysis, and the average current density is calculated from the total charge amount 30 seconds before each sampling. Quantitative analysis of gaseous products was calibrated with the standard gas mixtures from Shanghai Haizhou Special Gas Co., LTD.

**Acknowledgement**

We acknowledge the funding supports from the National Foundation of Natural Science, China (52022002). We thank Prof. Minhua Shao for the equipment support and Dr. Shangqian Zhu for assistance with characterizing CO$_2$ reduction.

# Supporting Information

Figure S1. A photo of the pycnometer measurement.

Figure S2. Morphologies of all NP metals examined in the work but not shown in Fig. 1.

Figure S3. A photo of the measurement jig in operation.

Table S1. A summary of the pore sizes, the porosities and the measured values of $\sigma_{eff}/\sigma_0$.

Figure S4. The temperature dependence of the coarsening kinetics of NP Au.

Figure S5. EIS curves with 0.5 M zinc acetate.

Figure S6. Measurements of the electrochemical areas of hierarchical and oriented NP Ag's.

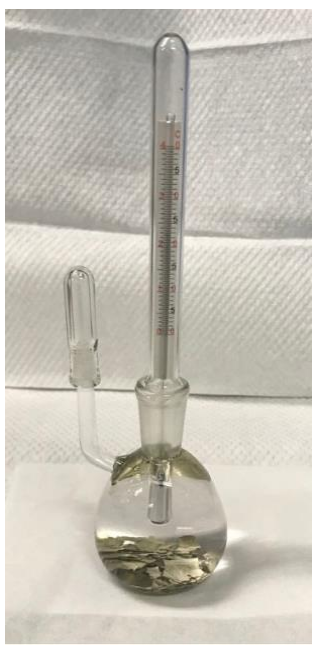

Figure S1. A photo of the pycnometer measurement. The bottle has a capacity of $V_0$. After placing NP metal in the bottle, we filled the bottle with liquid (density $\rho_1$ and a volume of $V_1$) until it reaches the full mark indicated in the side capillary. We then measured the weight of the bottle with the liquid and the NP metal ($M_1$), and the weight of the bottle filled with just the liquid ($M_2$). The volume of pores inaccessible by the liquid (*i.e.*, closed pores) is $V_0 - V_1 - [M_1 - M_2 + \rho_1(V_0 - V_1)]/\rho_2$, where $\rho_2$ is the density of the solid metal.

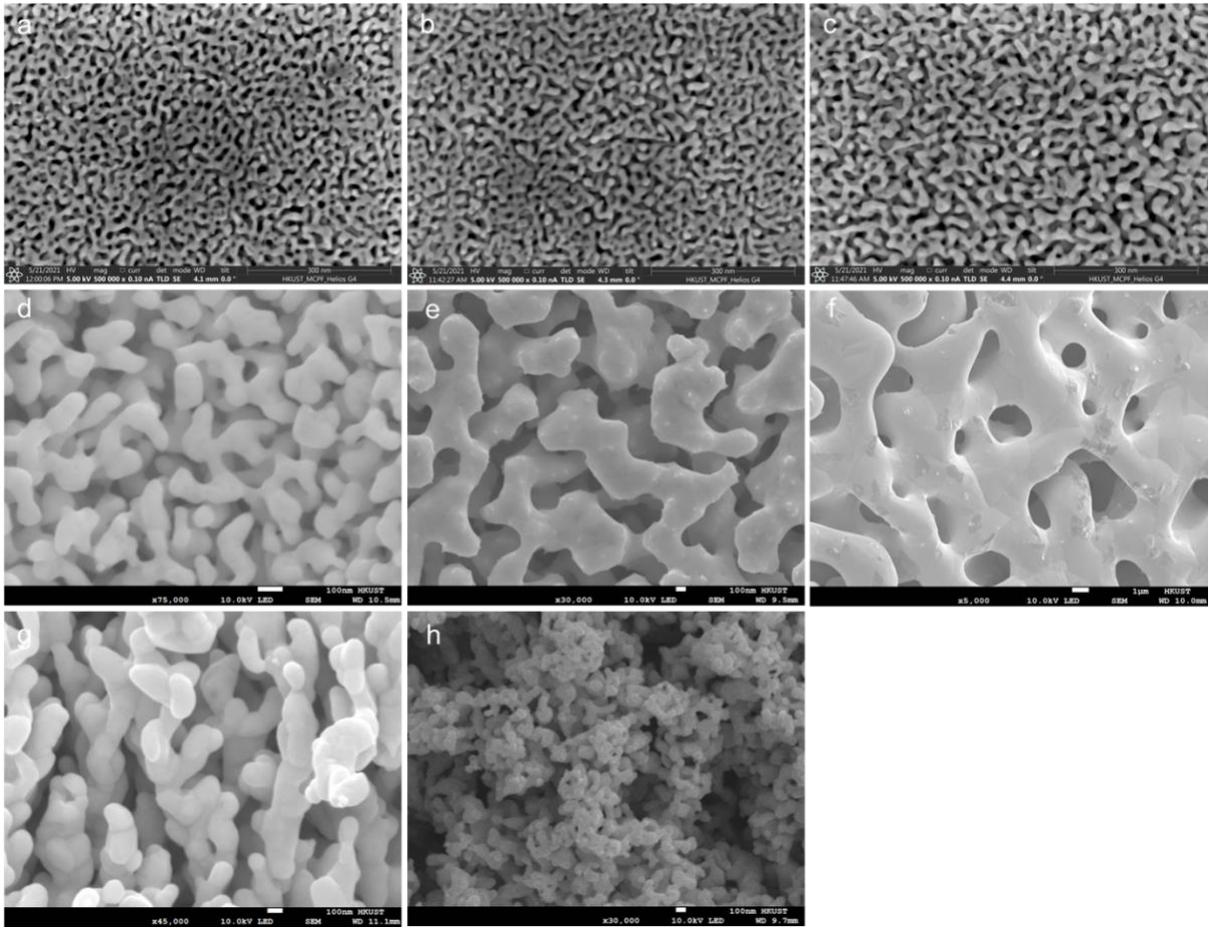

Figure S2. Morphologies of all NP metals examined in the work but not shown in Fig. 1. (a – f) NP Au coarsened at 100, 150, 200, 400, 800 ºC for 2 hours, respectively. (g) NP Ag coarsened at 100 ºC for 1 hour. (h) Hierarchical, low-$\varepsilon$ NP Ag fabricated via the RID of a sintered $Ag_2O$ powder precursor.

Table S1. A summary of the pore sizes, the porosities and the measured values of $\sigma_{eff}/\sigma_0$.

| Sample | Porosity, $\varepsilon$ | Pore size, $d$ (nm) | $\sigma_{eff}/\sigma_0$ |
|---|---|---|---|
| NP Au | 0.689 | 12.6 ± 4 | 0.1 |
| NP Au (50 °C) | 0.685 | 15.3 ± 4 | 0.21 |
| NP Au (100 °C) | 0.685 | 18.1 ± 4 | 0.38 |
| NP Au (150 °C) | 0.685 | 21.4 ± 4 | 0.41 |
| NP Au (200 °C) | 0.68 | 112 ± 26 | 0.43 |
| NP Au (400 °C) | 0.62 | 309 ± 94 | 0.37 |
| NP Au (600 °C) | 0.473 | 979 ± 302 | 0.26 |
| NP Au (800 °C) | 0.379 | 2447 ± 531 | 0.2 |
| NP Ag | 0.610 | 94.5 ± 28 | 0.18 |
| NP Ag (100 °C) | 0.601 | 228 ± 72 | 0.18 |
| NP Ag (200 °C) | 0.579 | 499 ± 156 | 0.17 |
| Low-$\varepsilon$ NP Ag (RID of Ag$_2$O) | 0.365 | 65 ± 22 | 0.07 |
| Hierarchical NP Ag (AgCl powder precursor) | 0.784 | 68.1 ± 22 | 0.44 |
| Hierarchical NP Ag (Ag$_2$O powder precursor) | 0.497 | 65 ± 22 | 0.17 |
| Oriented NP Ag | 0.544 | 94.5 ± 28 | 0.28 |

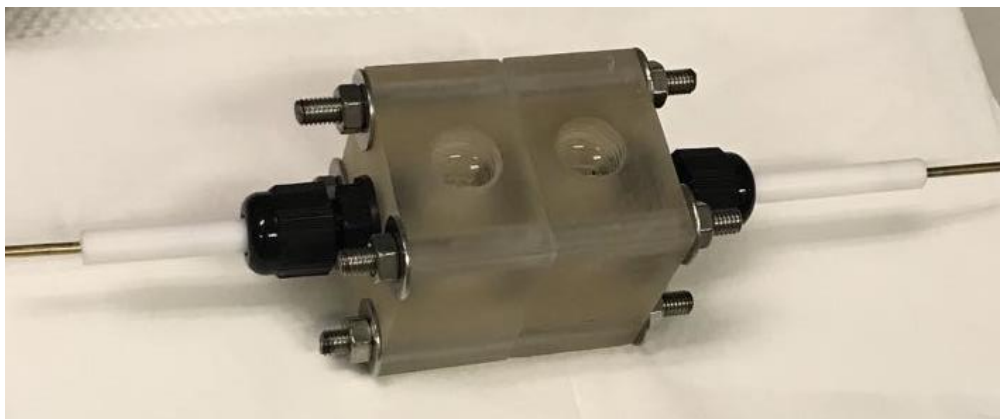

Figure S3. A photo of the measurement jig in operation.

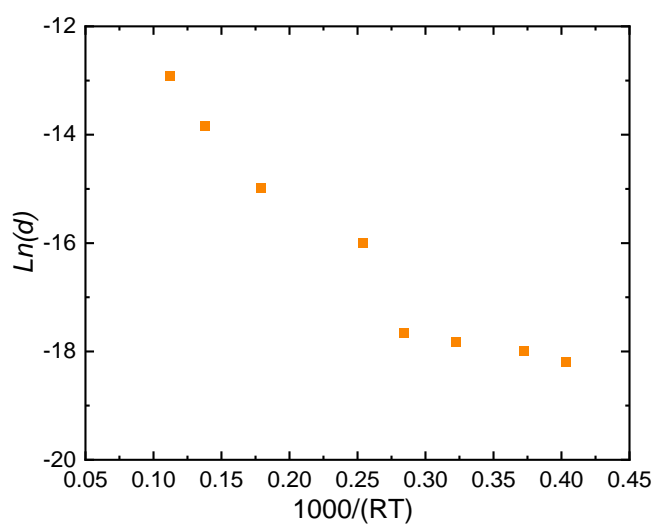

Figure S4. The temperature dependence of the coarsening kinetics of NP Au. The logarithm of $d$ is plot against 1000/(RT), where R is the gas constant and T the coarsening temperature in Kelvin.

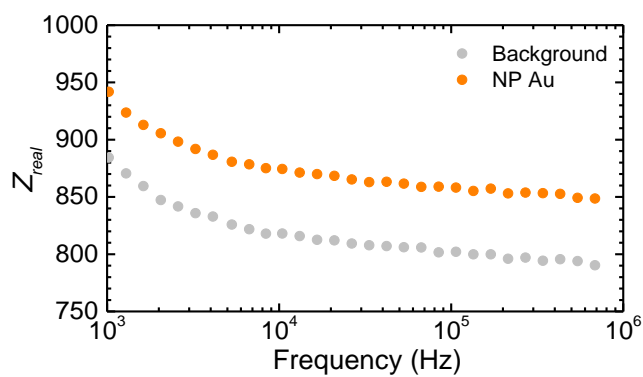

Figure S5. EIS curves of the background and NP Au ($d$ = 19 nm) measured with 0.5 M zinc acetate. The high frequency responses are not as stable likely because of the adsorption of the ions on the Pt probes, but ΔR remains stable in the range of 0.1 – 1 MHz.

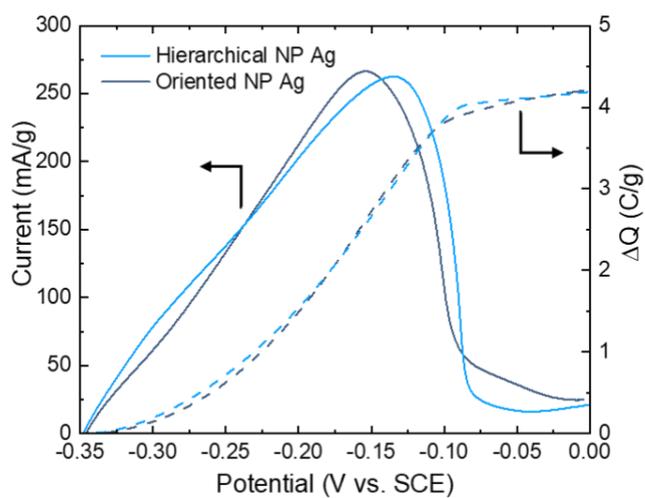

Figure S6. Measurements of the electrochemical areas of hierarchical and oriented NP Ag's via the stripping a layer of underpotential Pb deposit. The linear sweep voltammetry (solid lines) was scanned from the potential of deposition (-0.35 vs. the saturated calomel electrode, SCE) until the charge (dashed lines) reached a plateau. Details of the method can be found in Ref. 15. The estimated areas of the hierarchical and oriented structures are 6.9 and 7.0 m$^2$/g, respectively, close to that of NP Ag (7.2 m$^2$/g).